\documentclass[aps,prd,twocolumn,superscriptaddress,altaffilletter,nobibnotes,nofootinbib,10pt,showpacs,showkeys,preprintnumbers]{revtex4-2}

\usepackage{float}
\usepackage{graphicx}
\usepackage{dcolumn}
\usepackage[dvipsnames]{xcolor}
\usepackage{bm}
\usepackage{lineno}
\usepackage{float}
\usepackage{amsmath}
\usepackage{soul}
\usepackage{multirow}
\usepackage{tikz,xcolor,hyperref}

\definecolor{lime}{HTML}{A6CE39}
\DeclareRobustCommand{\orcidicon}{
	\begin{tikzpicture}
	\draw[lime, fill=lime] (0,0) 
	circle [radius=0.16] 
	node[white] {{\fontfamily{qag}\selectfont \tiny ID}};
	\draw[white, fill=white] (-0.0625,0.095) 
	circle [radius=0.007];
	\end{tikzpicture}
	\hspace{-2mm}
}
\foreach \x in {A, ..., Z}{\expandafter\xdef\csname orcid\x\endcsname{\noexpand\href{https://orcid.org/\csname orcidauthor\x\endcsname}
			{\noexpand\orcidicon}}
}

\newcommand{\RomanNumeralCaps}[1]
{\MakeUppercase{\romannumeral #1}}

\newcommand{\pte}{$p_{\rm{_T}}$}
\newcommand{\pt}{$p_{\rm{_T}}$~}

\newcommand{\pb}{$Pb-Pb$~}

\begin{document}


\title{A unified formalism to study the pseudorapidity spectra in heavy-ion collision}
\author{Rohit Gupta}
\affiliation{Department of Physical Sciences, Indian Institute of Science Education and Research (IISER) Mohali, Sector 81 SAS Nagar, Manauli PO 140306 Punjab, India}

\author{Aman Singh Katariya}%
\affiliation{Department of Physical Sciences, Indian Institute of Science Education and Research (IISER) Mohali, Sector 81 SAS Nagar, Manauli PO 140306 Punjab, India}
\author{Satyajit Jena}
\email{sjena@iisermohali.ac.in}
\affiliation{Department of Physical Sciences, Indian Institute of Science Education and Research (IISER) Mohali, Sector 81 SAS Nagar, Manauli PO 140306 Punjab, India}

\begin{abstract}
The pseudorapidity distribution of charged hadron over a wide $\eta$ range gives us crucial information about the dynamics of particle production. The constraint by the detector acceptance, particularly at forward rapidities, demands a proper distribution function to extrapolate the pseudorapidity distribution to large $\eta$. In this work, we have proposed a phenomenological model based on the Pearson statistical framework to study the pseudorapidity distribution. We have analyzed and fit data of charged hadrons produced in \pb collision at $2.76$ TeV  and $Xe-Xe$ collision at $5.44$ TeV using the proposed model.  

\end{abstract}

\maketitle

\section{\label{sec:intro}INTRODUCTION}
Evidence such as the quarkonia suppression, jet quenching, and strangeness enhancement points toward the formation of a new state of hot and dense nuclear matter under the extreme condition of temperature and energy density. This state of asymptotically free quarks \& gluons is popularly known as Quark-Gluon Plasma (QGP). Characterizing the QGP, understanding the confinement-deconfinement phase transition and the search for QCD critical point are some of the open problems that drive the bulk of the heavy-ion physics research. 

The short time scale of the QGP formation ($\sim$ few fm/c) and threshold in the detector capabilities limit the direct measurement of QGP.  However, the kinematic observables of the final state particles that are streaming to the detectors bring sufficient information to characterize QGP, and hence, act as a standard candle in the studies related to the evolution of the system created during high energy collision.

Transverse momentum (\pte), pseudorapidity ($\eta$), energy ($E$) and azimuthal angle ($\phi$) are some of the kinematic observables  used to characterize a track in the experiment. Pseudorapidity distribution $(\eta = -ln(tan(\theta/2)))$ of charged particles can provide us important insight into the initial energy density and the mechanism of particle production in heavy-ion collision.  The collision system and centrality dependence of the pseudorapidity distribution of charged particles over a wide $\eta$ range are crucial to understand the relative contribution of hard scattering processes and soft processes in particle production \cite{Adam:2016ddh}.

Measurement of the kinematics observable such as \pt and $\eta$ of charged hadrons is limited to the mid-rapidity region because of the complexity of introducing the detector at forward rapidities. Traditionally, detectors are installed in the form of cylindrical geometry, mostly covering the central rapidity regions. Hence, a theoretical model that could extrapolate the pseudorapidity distribution with good precision is of interest to understand the dynamics of particle production.

Many theoretical studies have been performed in this direction to study the pseudorapidity distribution. A Multi-source thermal model and its Tsallis generalization have been discussed in Ref.~\cite{Sun:2013ota, Li:2014opa}. This model assumes four sources, target and projectile cylinder and target and projectile leading particles. A three source scenario for particle production within the non-equilibrium statistical relativistic diffusion model has been explored in Ref.~\cite{Wolschin:2011mz}. It has been argued that the midrapidity source corresponds to the charged hadrons produced in gluon-gluon interaction, whereas the other two sources at large rapidities correspond to the quark-gluon interactions.  A large fraction of charged hadrons belongs to the midrapidity source, and their relative particle content controls the size of the dip in midrapidity.  Although the Jacobian transformation from $y$ space to $\eta$ space does not significantly influence the midrapidity dip \cite{Wolschin:2011mz}, the dip at midrapidity is a mathematical artifact of particle production at $\theta = 90^o$. Another variant of the three source scenario has been discussed in Ref.~\cite{Gao:2015sdb} where the central source of the particle is described using the Landau hydrodynamics model.  Several other models have also been developed to study the pseudorapidity distribution. 

The Boltzmann statistics have been used in a scenario with a large number of fireballs to study the pseudorapidity distribution of charged hadrons produced in $pp$ and heavy-ion collision \cite{Cleymans:2008zz, Becattini:2007qr}.  Tsallis generalization for the same is discussed in the Ref.~\cite{Marques:2015mwa, Gao:2017yas, Tao:2020uzw}. The Tsallis formalism assumes two clusters of fireballs and the rapidity distribution are given in terms of $q$-Gaussian distribution.  However, all these approaches do not consider hard processes, which is also a part of the particle production mechanism. Recently, a generalization of Tsallis distribution, providing a consistent description of particle production in soft processes and the hard-scattering processes, has been discussed in Ref.~\cite{Jena:2020wno, Gupta:2020naz}. In this work, we present the application of the Pearson statistical framework to study the pseudorapidity distribution of charged hadrons produced in \pb collision at $2.76$ TeV \cite{Abbas:2013bpa, Adam:2015kda}  and $Xe-Xe$ collision at $5.44$ TeV  \cite{Acharya:2018hhy} measured by the ALICE experiment. 

We will start with the discussion of  theoretical formalism based on the Pearson statistical framework to study the pseudorapidity distribution. The section \RomanNumeralCaps{3} will cover the results obtained by fitting the charged hadron spectra at two different energies, followed by the conclusion.

\section{Theoretical Description}

Pearson distribution is a generalized distribution function introduced by Karl Pearson \cite{Pearson343} in 1895. It is characterized using the first four moments related to the mean, standard deviation, skewness and kurtosis of a distribution. Pearson function is described in terms of a differential equation \cite{pollard}:
\begin{equation}\label{diffe}
 \frac{1}{p(x)}\frac{dp(x)}{dx} + \frac{a + x}{b_0 + b_1 x + b_2 x^2} = 0
\end{equation}
with the corresponding parameters a, $b_0$, $b_1$ and $b_2$ given in term of the first four central moments $m_1, m_2, m_3$ and $m_4$ as:
\begin{equation}
a = b_1 = \frac{m_3(m_4 + 3 m_2^2)}{10m_2m_4 - 18 m_2^3 - 12 m_3^2}
\end{equation}
\begin{equation}
b_0 = \frac{m_2(4m_2m_4 - 3 m_3^2)}{10m_2m_4 - 18 m_2^3 - 12 m_3^2}
\end{equation}
\begin{equation}
b_2 = \frac{2m_2m_4 - 6 m_2^3 - 3 m_3^2}{10m_2m_4 - 18 m_2^3 - 12 m_3^2}
\end{equation}
In some of the recent works \cite{Jena:2020wno, Gupta:2020naz, Gupta:2021olm}, the Pearson statistical framework has been described to study the distribution of particles produced in hadron-hadron and heavy-ion collision at different energies. Pearson formalism has been discussed as an extension to Tsallis statistics to explain the particle production in soft processes as well as hard scattering processes in a consistent manner.  Pearson formalism is thermodynamically consistent, and within some limit on its parameters, it reduces to the Tsallis distribution upto some normalization constant. Energy distribution of final state particles in this formalism is given as:
\begin{equation}
E \frac{d^3 N}{dp^3} = B'  \left( 1 + \frac{E}{p_0}\right) ^{-n} \left(  1 + (q-1)\frac{(E-\mu)}{T}\right)^{-\frac{q}{q-1}}
\end{equation}
Replacing $E$ with $m_Tcosh~y$ and setting $\mu=0$ at the LHC energies, the equation above gets simplified to:
\begin{equation}\label{pearsonpt}
\begin{aligned}
 \frac{1}{2\pi p_T} \frac{d^2 N}{dp_T dy} = B' & \left( 1 + \frac{m_Tcosh~y}{p_0}\right) ^{-n}\\
&\left(  1 + (q-1)\frac{m_Tcosh~y}{T}\right)^{-\frac{q}{q-1}}
\end{aligned}
\end{equation}
Considering the suggestions given in Ref.~\cite{Marques:2015mwa},  we assume that clusters of fireballs travel with ultra-relativistic velocity and consists of several small fireballs. Considering a fireball moving in the laboratory frame with the rapidity $y_f$, the rapidity distribution of corresponding secondaries produced from the decay of the fireball can be obtained by integrating the above equation over \pte:
\begin{equation}
\begin{aligned}
     \frac{dN}{dy} = A  \int_0^{\infty} & dp_T p_T\left( 1 + \frac{m_Tcosh(y-y_f)}{p_0}\right) ^{-n}\\
&\left(  1 + (q-1)\frac{m_Tcosh(y-y_f)}{T}\right)^{-\frac{q}{q-1}}
\end{aligned}
\end{equation}
We will also consider a large number of such fireballs distributed over the rapidity space with the distribution $\nu(y_f)$ given in term of a double q-Gaussian function as:
\begin{equation}\label{vyf_init}
    \nu \left(y_{f}\right) = G(y_0,\sigma;y_f) + G(-y_0,\sigma;y_f)
\end{equation}
Here $G(y_0,\sigma;y_f)$ is the q-Gaussian given as:
\begin{equation}
    G(y_0,\sigma;y_f) = \frac{1}{\sqrt{2\pi}\sigma}e_q\left( -\frac{(y_f-y_0)^2}{2\sigma^2} \right)
\end{equation}
and the q-exponential $e_q(x)$ \cite{Tsallis:1987eu} is defined as:
\begin{equation}
    e_q(x) \equiv [1-(q-1)x]^{-\frac{1}{q-1}}
\end{equation}
On substituting q-exponential and q-Gaussian in Eq.~(\ref{vyf_init}) we get:
\begin{equation}\label{dist_fireball}
\begin{aligned}
    \nu \left(y_{f}\right) =& \frac{1}{\sqrt{2 \pi} \sigma} \left[ 1+\left(q -1 \right) \frac{\left(y_{f}-y_{0}\right)^2}{2 \sigma^2}\right]^{-\frac{1}{q-1}}\\
    & +\frac{1}{\sqrt{2 \pi} \sigma} \left[ 1+\left(q -1 \right) \frac{\left(y_{f}+y_{0}\right)^2}{2 \sigma^2}\right]^{-\frac{1}{q-1}}
\end{aligned}
\end{equation}
In Eq.~(\ref{dist_fireball}), $y_0$ represents the peak position and $\sigma$ is the width of the q-Gaussian.

The formalism developed here is expressed as a function of rapidity $y$; however, most of the experimental results are presented as a function pseudorapidity $\eta$ as always mass of the particle is not known. Thus, the Jacobian transformation relating $dy$ and $d\eta$ can be used and given as:
\begin{equation}
    \frac{dy}{d\eta} = \sqrt{1-\frac{m_0^2}{m_T^2 cosh~y}}
\end{equation}
Further, the relation between y and $\eta$ will be of the form:
\begin{equation}\label{rap_pseudo}
    y = \frac{1}{2} \ln \left[ \frac{\sqrt{p_T^2 \cosh^2 \eta + m_0^2}+p_T\sinh \eta}{\sqrt{p_T^2 \cosh^2 \eta + m_0^2}-p_T\sinh \eta}\right]
\end{equation}
Finally, the distribution of charged hadrons pseudorapidity can be obtained by integrating over \pt and $y_f$ in the equation:
\begin{equation}
\begin{aligned} \label{pseudorap}
\frac{d N}{d \eta}=& A \int_{-\infty}^{\infty} d y_{f} \int_{0}^{\infty} d p_{T} p_{T} \sqrt{1-\frac{m_{0}^{2}}{m_{T}^{2} \cosh ^{2} y}} \\
& \times \nu\left(y_{f}\right)  \left( 1 + \frac{m_Tcosh(y-y_f)}{p_0}\right) ^{-n}\\
& \times \left[1+(q-1) \frac{m_{T} \cosh \left(y-y_{f}\right)}{T}\right]^{-\frac{q}{q-1}}
\end{aligned}
\end{equation}
with the distribution of fireball represented in Eq.~(\ref{dist_fireball}) and the rapidity $y$ will be replaced by Eq.~(\ref{rap_pseudo}). It is not possible to analytically perform this integral; hence, we have carried out the numerical integration to evaluate the integral. We have used Eq.~(\ref{pseudorap}) to fit the pseudorapidity distribution of charged hadron with parameters $y_0$ and $\sigma$ considered as the free parameters that need to be adjusted to obtain the best fit to the data. The rest of the parameters $q, T, p_0$ and $n$ are obtained by fitting the corresponding \pte-spectra with the Pearson distribution function as shown in Eq.~(\ref{pearsonpt}). Once we obtain the distribution function, the total number of charged particles can be estimated by integrating the above equation over the range of pseudorapidity.
\begin{figure}[h!]
  \includegraphics[height=2in,width=3in]{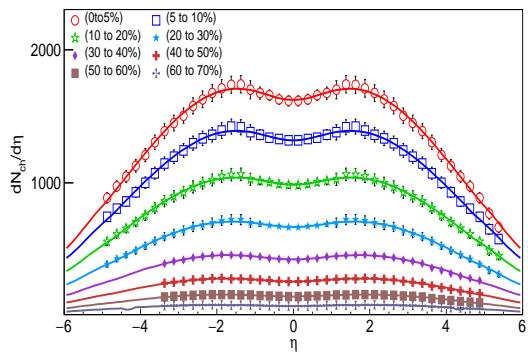}
  \caption{(color online) The charged hadron pseudorapidity distribution for $\sqrt{s_{NN}} = 2.76$ TeV \pb collision, fitted with Pearson pseudorapidity distribution over the range $-6 \leq \eta \leq 6$.}\label{fig:2760pearson}
\end{figure}
\begin{figure}[h!]
  \includegraphics[height=2in,width=3in]{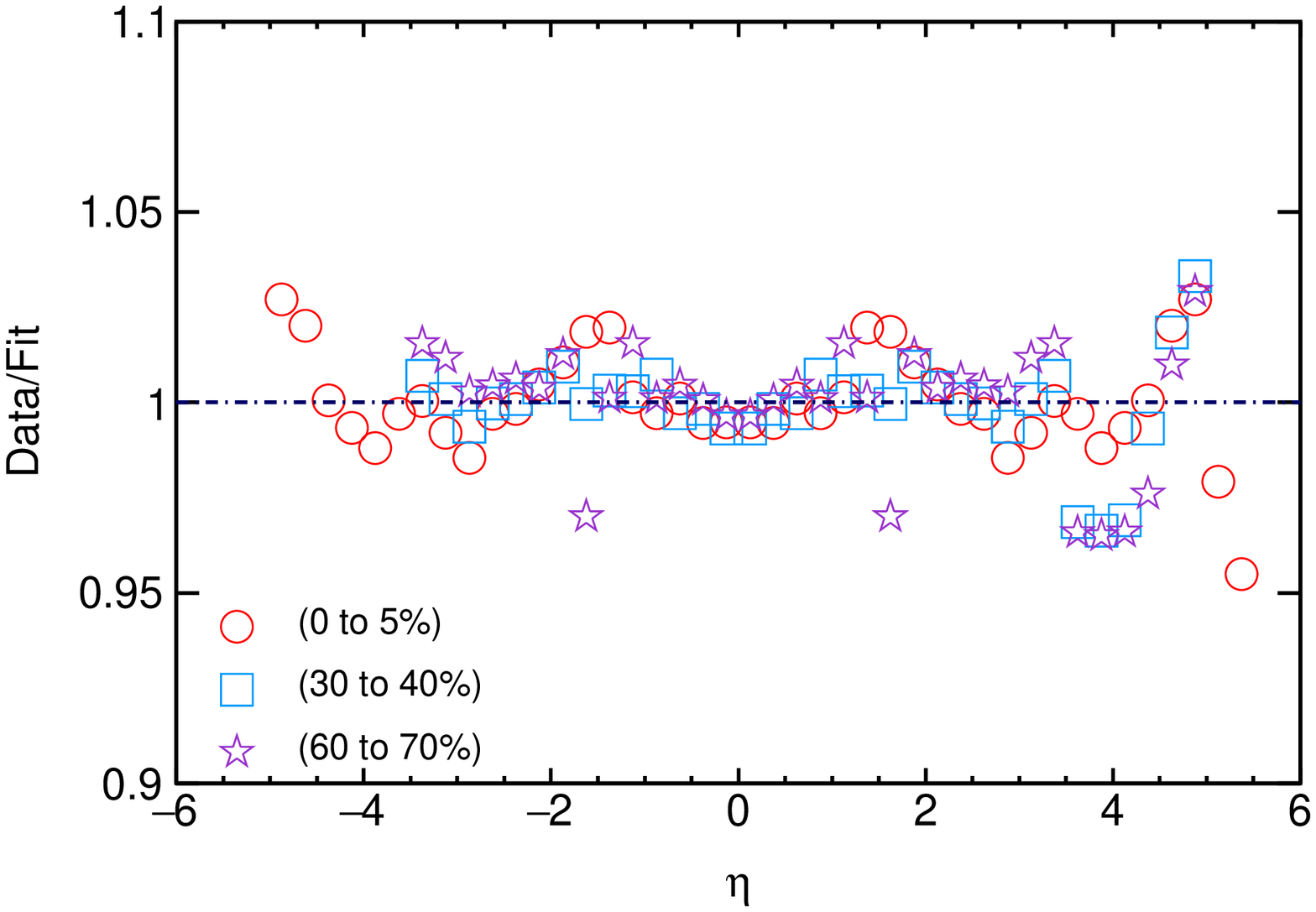}
  \caption{(color online) Ratio of data to the fit function for three different centralities of \pb collision at 2.76 TeV.}\label{fig:ratio_2760}
\end{figure}

\begin{table}
\centering
\caption{\label{tab:chi}The $\chi^{2}/NDF$ values for the pseudorapidity data at $2.76$ TeV and $5.44$ TeV fitted with the distribution function Eq.~(\ref{pseudorap}).}
\begin{tabular*}{\columnwidth}{@{\extracolsep{\fill}}lll@{}}
\hline
\multirow{2}{*}{Centrality} & \multicolumn{2}{c}{$\chi^2/NDF$} 
\\ 
\cline{2-3}
 & $2.76$ TeV & $5.44$ TeV \\
\hline
0 to 5 \% & $3.99/39$ & - \\
5 to 10 \% & $3.135/39$ & - \\
10 to 20 \% & $2.757/39$ & $3.671/53$ \\
20 to 30 \% & $1.988/39$ &  $3.366/53$ \\
30 to 40 \% & $1.611/31$ &  $2.831/53$ \\
40 to 50 \% & $1.407/31$ &  $3.5/53$  \\
50 to 60 \% & $1.269/31$ &  $5.112/53$ \\
60 to 70 \% & $4.557/31$ & $4.849/53$  \\
70 to 80 \% & - & $4.382/53$ \\
\hline
\end{tabular*}
  
\end{table}

\section{Results and Discussion}
The aforementioned formalism is tested and verified on the experimental data of $dN_{ch}/d\eta$ distribution of charged hadron produced in $2.76$ TeV \pb
\cite{Abbas:2013bpa, Adam:2015kda} and $5.44$ TeV $Xe-Xe$ \cite{Acharya:2018hhy} collision.
As discussed earlier, this model needs the thermodynamical parameters such as $q, T, p_0$ and $n$. These parameters are not directly measurable quantities and hence needed to be extracted by fitting to the respective \pt-spectra. In order to extract these parameters, the data of charged hadrons transverse momentum spectra measured in $2.76$ TeV \pb \cite{Abelev:2012hxa} and $5.44$ TeV $Xe-Xe$ \cite{Acharya:2018eaq} is fitted with the Pearson distribution function.

%
We have implemented our formalism in the ROOT \cite{root} data analysis framework and the MINUIT \cite{James:1975dr} algorithm is used to obtain best fit results. We have tested the goodness of fit using the $\chi^2/NDF$ values.

\begin{table}
\centering
\caption{\label{tab:pbpb}Numerical values of $dN_{ch}/d\eta$ measured by the experiments for \pb collision at $2.76$ TeV along with the values obtained from the fit function Eq.~(\ref{pseudorap}) and the ratio (data/fit) for two different centralities.}
\begin{tabular*}{\columnwidth}{@{\extracolsep{\fill}}lllllll@{}}
\hline
\multirow{2}{*}{$\eta$} & \multicolumn{3}{c|}{\textbf{Centrality 0 to 5 \%}} &  
\multicolumn{3}{c}{\textbf{Centrality 60 to 70 \%}}
\\ 
\cline{2-7}
 & Data & Fit& Ratio & Data & Fit & Ratio \\
\hline
-3.375 & 1388 & 1387.43 & 1.00 &  76.10 & 74.94& 1.01\\
-2.875 & 1504 & 1526.06 & 0.98 & 80.00 & 79.77& 1.00\\
-2.375 & 1627 & 1629.76 & 1.00 & 83.10 & 82.58 & 1.01\\
-1.875 & 1709 & 1691.15 & 1.01 & 84.30 & 83.24 & 1.01\\
-1.375 & 1739 & 1705.59 & 1.02 & 81.90 & 81.81 & 1.00\\
-0.875 & 1674 & 1678.64 & 1.00 & 78.90 & 78.82 & 1.00\\
-0.375 & 1627 & 1635.87 & 0.99 & 75.80 & 75.76 & 1.00\\
0.375 & 1627 & 1635.87 & 0.99 & 75.80 & 75.76 & 1.00\\
0.875 & 1674 & 1678.64 & 1.00 & 78.90 & 78.82 & 1.00\\
1.375 & 1739 & 1705.59 & 1.02 & 81.90 & 81.81 & 1.00\\
1.875 & 1709 & 1691.15 & 1.01 & 84.30 & 83.24 & 1.01\\
2.375 & 1627 & 1629.76 & 1.00 & 83.10 & 82.58 & 1.00\\
2.875 & 1504 & 1526.06 & 0.98 & 80.00 & 79.77 & 1.00\\
3.375 & 1388 & 1387.43 & 1.00 & 76.10 & 74.94 & 1.02\\
3.875 & 1209 & 1223.56 & 0.99 & 66.00 & 68.38 & 0.96\\
4.375 & 1046 & 1045.42 & 1.00 & 59.10 & 60.55 & 0.98\\
4.875 & 888 & 864.677 & 1.03 & 53.60 & 52.08 & 1.03\\
\hline
 \end{tabular*}
\end{table}

\begin{table}
\centering
\caption{\label{tab:XeXe}Numerical values of $dN_{ch}/d\eta$ measured by the experiments for $Xe-Xe$ collision at $5.44$ TeV along with the values obtained from the fit function Eq.~(\ref{pseudorap}) and the ratio (data/fit) for two different centralities.}
\begin{tabular*}{\columnwidth}{@{\extracolsep{\fill}}lllllll@{}}
\hline
\multirow{2}{*}{$\eta$} & \multicolumn{3}{c|}{\textbf{Centrality 10 to 20 \%}} &  
\multicolumn{3}{c}{\textbf{Centrality 70 to 80 \%}}
\\ 
\cline{2-7}
& Data & Fit& Ratio & Data & Fit & Ratio \\
\hline
-3.375 & 617.80 & 621.20 & 0.99 & 31.70 & 31.35 & 1.01\\
-2.875 & 669.90 & 670.69 & 1.00 & 33.50 & 32.76 & 1.02\\
-2.375 & 696.10 & 704.72 & 0.99 & 33.40 & 33.74 & 0.99\\
-1.75 & 732.20 & 731.57 & 1.00 & 33.60 & 34.22 & 0.98\\
-1.25 & 741.30 & 735.22 & 1.00 & 34.00 & 33.84 & 1.00\\
-0.75 & 725.50 & 724.08 & 1.00 & 33.00 & 32.89 & 1.00\\
-0.25 & 704.90 & 709.88 & 0.99 & 31.90 & 31.99 & 1.00\\
0.25 & 704.90 & 709.88 & 0.99 & 31.90& 31.99 & 1.00\\
0.75 & 725.50 & 724.08 & 1.00 & 33.00 & 32.89 & 1.00\\
1.25 & 741.30 & 735.22 & 1.00 & 34.00 & 33.84 & 1.00\\
1.75 & 732.20 & 731.57 & 1.00 & 33.60 & 34.22 & 0.98\\
2.375 & 696.10 & 704.72 & 0.99 & 33.40 & 33.74 & 0.99\\
2.875 & 669.90 & 670.69 & 1.00 & 33.50 & 32.76 & 1.02\\
3.375 & 617.80 & 621.20 & 0.99 & 31.70 & 31.35 & 1.01\\
3.875 & 547.10 & 576.31 & 0.95 & 28.20 & 29.63 & 0.95\\
4.375 & 520.10 & 512.91 & 1.01 & 27.60 & 27.64 & 1.00\\
4.875 & 473.30 & 459.63 & 1.03 & 25.80 & 25.58 & 1.01\\
\hline
 \end{tabular*}
\end{table}

\begin{figure}[h!]
  \includegraphics[height=2in,width=3in]{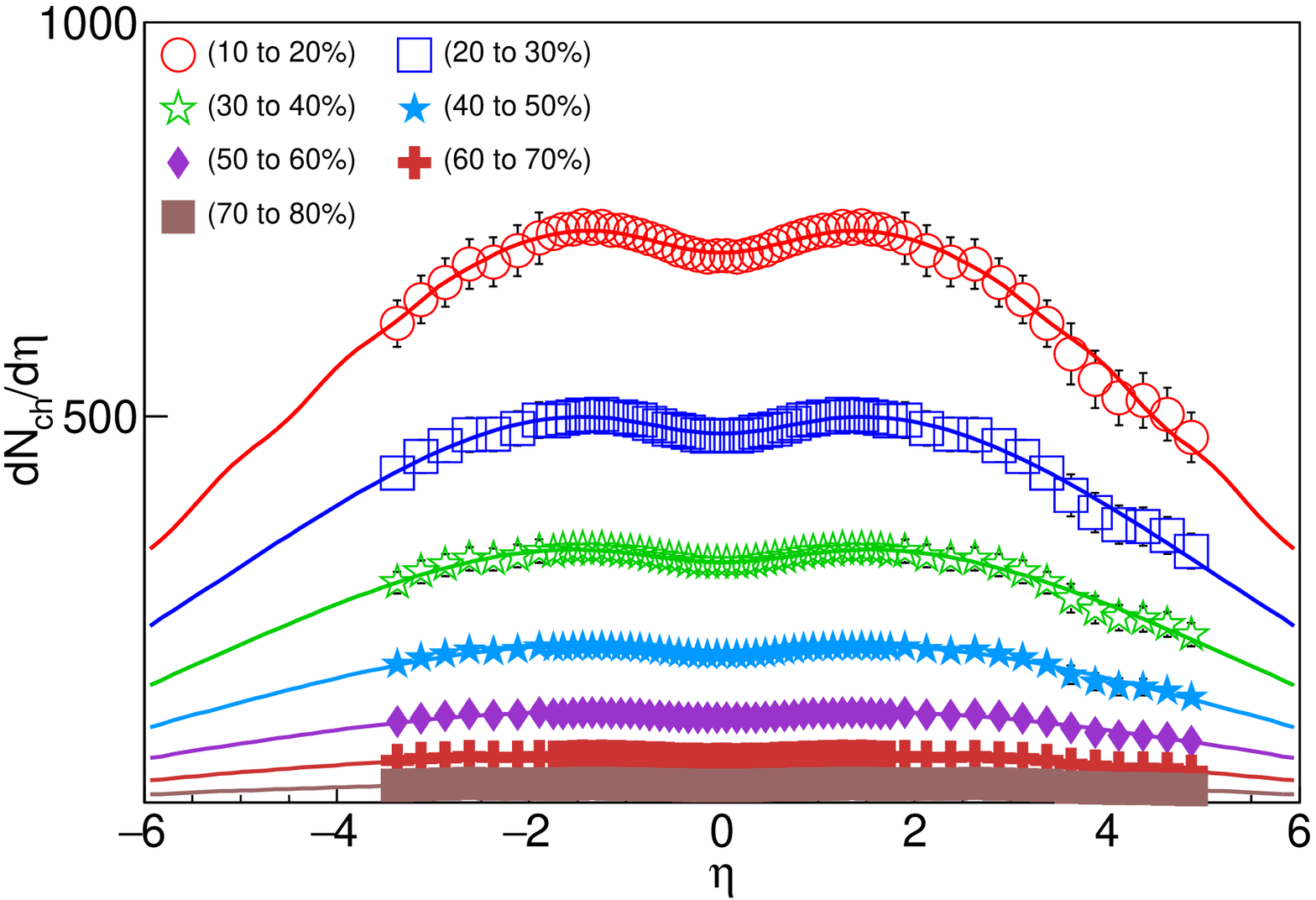}
  \caption{(color online) The charged hadron pseudorapidity distribution for $\sqrt{s_{NN}} = 5.44$ TeV  $Xe-Xe$ collision, fitted with Pearson pseudorapidity distribution over the range $-6 \leq \eta \leq 6$.}\label{fig:5440pearson}
\end{figure}
\begin{figure}[h!]
  \includegraphics[height=2in,width=3in]{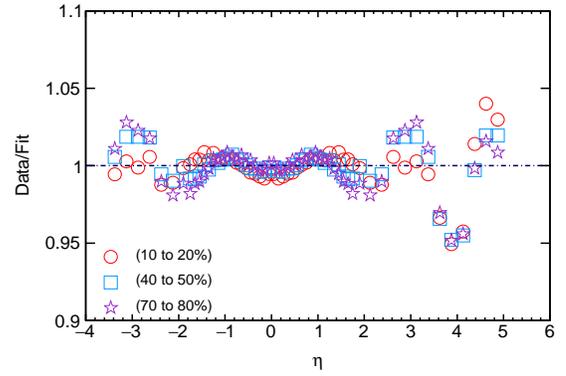}
  \caption{(color online) Ratio of data to the fit function obtained for three different centralities $Xe-Xe$ collision data at 5.44 TeV.}\label{fig:ratio_5440}
\end{figure}

\begin{figure}[h!]
  \includegraphics[height=2in,width=3in]{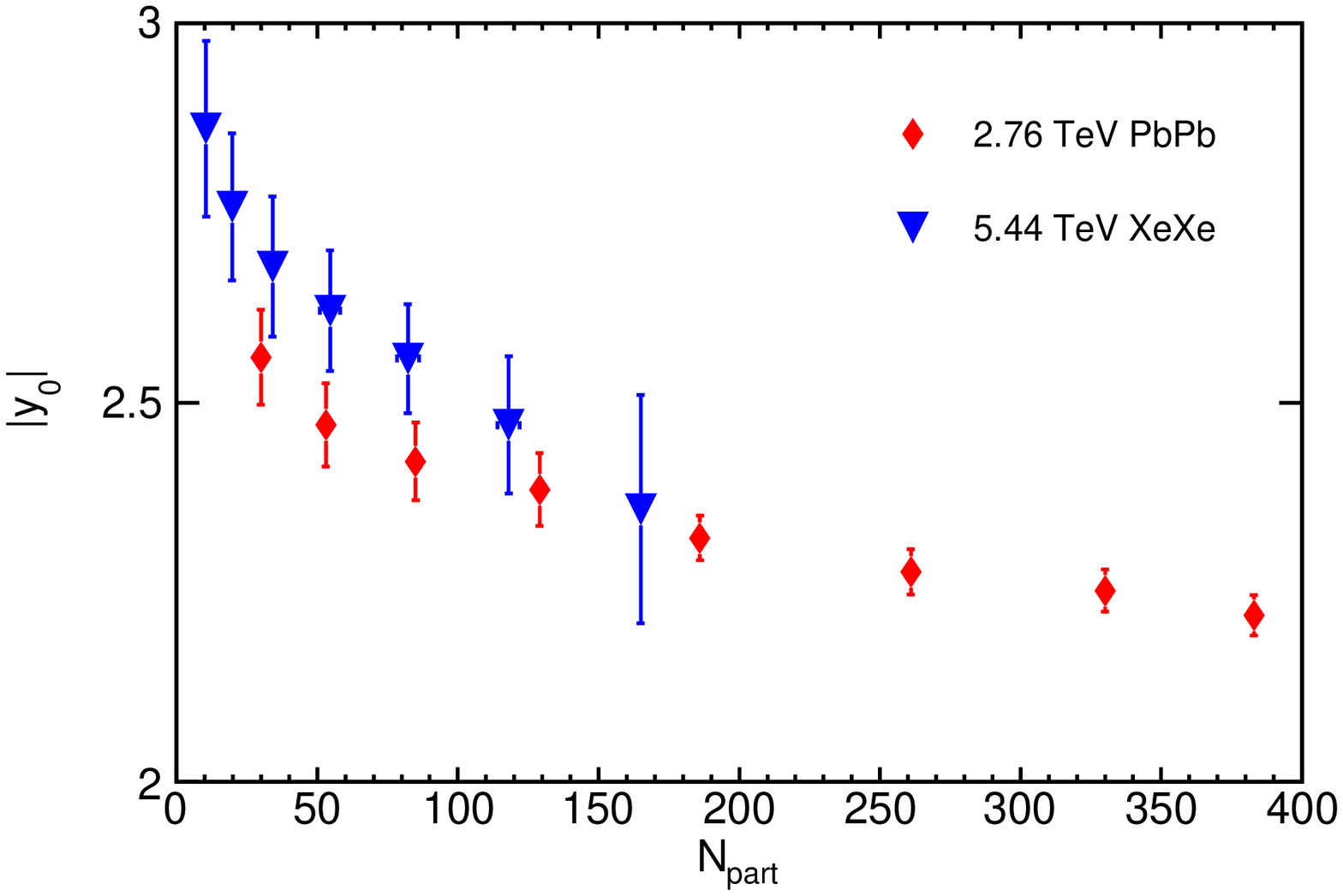}
  \caption{(color online) Variation of parameter $|y_0|$ with the charged particle multiplicity for two different energies.}\label{fig:y0}
\end{figure}

\begin{figure}[h!]
  \includegraphics[height=2in,width=3in]{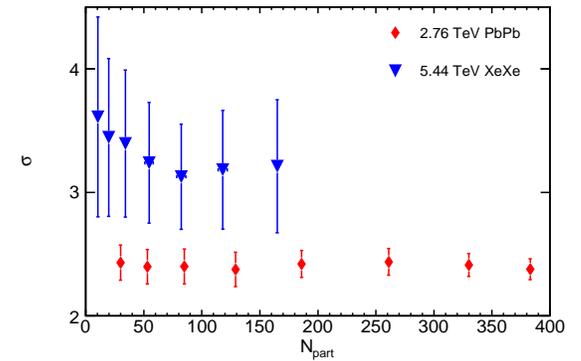}
  \caption{(color online) Variation of parameter $\sigma$ with the charged particle multiplicity for two different energies.}\label{fig:sig}
\end{figure}

The Fig.~\ref{fig:2760pearson}, \ref{fig:5440pearson} present the different centrality charged hadron pseudorapidity distribution data at two different energies fitted with the Pearson pseudorapidity function Eq.~(\ref{pseudorap}) over a broad $\eta$ range. A nice agreement between data and corresponding fit function has been obtained across different centrality and collision energy. Low $\chi^2/NDF$ values (as can be verified from table \ref{tab:chi}) suggest a good agreement between the fit function and the corresponding data. This can be further verified from the tables \ref{tab:pbpb}, \ref{tab:XeXe} where we have presented the values obtained for the data and the corresponding fit function for two different energies.

From the value of $\chi^2/NDF$ presented in table \ref{tab:chi}, we observe that the best fit values are close to zero instead of unity which is expected based on the $\chi^2$ analysis. The $\chi^2$ is defined as:
 \begin{equation}\label{eq:chi2}
     \chi^2 = \sum_{i=1}^{N}\frac{(x_i - x_{ti})^2}{\sigma_i^2}
 \end{equation}
 where data is represented by the $x_i \pm \sigma_i$ and the value obtained from the theoretical model is $x_{ti}$. Since the same measurement is repeated a large number of times (in high energy collision, we study a large number of events), for a particular data point $i$, we will have a distribution of $x_i$ with its mean given as $\mu_i$ and standard deviation represented by $\sigma_i$. If the measurement is repeated $N_i$ number of times, and the index $j$ represent one particular trial with $0<j\leq N_i$, then the variance is $\sigma_i^2  = \frac{1}{N_i}\sum_{j=1}^{N_i}{(x_{ij} - \mu_{i})^2}$. Considering the theory is correct, $x_{ti}$ will represent the mean value of the distribution of $x_i$. So, if we make a distribution of $i^{th}$ term in the Eq.~(\ref{eq:chi2}), we obtain a distribution with its peak around one. Hence, each term in the summation contribute a value close to unity, so, in general, the value of $\chi^2/NDF$ is close to unity. However, the data of \pt and $\eta$ in the experiments is provided in term of mean and the standard deviation, \textit{i.e.}, $x_i$ itself represent the mean value. Hence, if the model is correct, each term in the summation contribute a value close to zero and the $\chi^2/NDF$ will be close to zero for the most accurate theoretical model.

In the multiple fireball scenario, we have considered that two clusters of the fireball are moving toward each other with the rapidity presented in Fig.~\ref{fig:y0} and the width of the $q-Gaussian$ representing the distribution of fireball rapidity within a cluster in Fig.~\ref{fig:sig}. Values of $|y_0|$ decrease with an increase in the number of participants in the collision.

The ratio plots Fig.~\ref{fig:ratio_2760}, \ref{fig:ratio_5440} show a value close to unity with an uncertainty of maximum $5\%$ over a broad $\eta$ range and different centrality bins. The goodness of fit obtained from the function Eq.~(\ref{pseudorap}) is remarkably accurate. 
From the results presented in this section, we can say that the formalism discussed above is a good description of the pseudorapidity distribution and  it reproduces the experimental data quite well. The natural extension of this work is to test the formalism for different energies starting from few GeV collision in RHIC energies upto the high energy LHC collision which may be released in our future works.

\section{Conclusion}

We have discussed the importance of the Pearson statistical framework in studying the pseudorapidity distribution of charged hadrons produced in high energy heavy-ion collision. A multiple fireball scenario has been considered to develop a distribution function to explain pseudorapidity data within the Pearson framework. We have used this model to analyze the data of the pseudorapidity distribution of charged hadrons produced in $2.76$ TeV \pb and $5.44$ TeV $Xe-Xe$ collision as measured by the ALICE experiment. The Pearson formalism reproduces the experimental data quite well. A good agreement with the data suggests that this formalism can be utilized to extrapolate the pseudorapidity distribution to large $\eta$ values beyond the detector acceptance required to gain more insight into the dynamics of particle production. This work further extends the usage of the Pearson statistical framework to the pseudorapidity space. 

 \section{Acknowledgement}
  R. Gupta would like to acknowledge the financial support provided by CSIR through fellowship number 09/947 (0067) 2015-EMR-1.
 

%


\begin{thebibliography}{9}
\bibitem{Adam:2016ddh}
J.~Adam \textit{et al.} [ALICE],
Phys. Lett. B \textbf{772}, 567-577 (2017)
[arXiv:1612.08966 [nucl-ex]].

\bibitem{Sun:2013ota}
J.~X.~Sun, C.~X.~Tian, E.~Q.~Wang and F.~H.~Liu,
Chin. Phys. Lett. \textbf{30}, 022501 (2013)
doi:10.1088/0256-307X/30/2/022501

\bibitem{Li:2014opa}
B.~C.~Li, Y.~Z.~Wang, F.~H.~Liu, X.~J.~Wen and Y.~E.~Dong,
Phys. Rev. D \textbf{89}, no.5, 054014 (2014)
doi:10.1103/PhysRevD.89.054014
[arXiv:1403.4025 [hep-ph]].

\bibitem{Wolschin:2011mz}
G.~Wolschin,
EPL \textbf{95}, no.6, 61001 (2011)
doi:10.1209/0295-5075/95/61001
[arXiv:1106.3636 [hep-ph]].

\bibitem{Gao:2015sdb}
L.~N.~Gao and F.~H.~Liu,
Adv. High Energy Phys. \textbf{2015}, 184713 (2015)
doi:10.1155/2015/184713
[arXiv:1509.08603 [nucl-ex]].

\bibitem{Cleymans:2008zz}
J.~Cleymans,
J. Phys. G \textbf{35}, 044017 (2008)
doi:10.1088/0954-3899/35/4/044017

\bibitem{Becattini:2007qr}
F.~Becattini and J.~Cleymans,
J. Phys. G \textbf{34}, S959-964 (2007)
doi:10.1088/0954-3899/34/8/S135
[arXiv:hep-ph/0701029 [hep-ph]].

\bibitem{Marques:2015mwa}
L.~Marques, J.~Cleymans and A.~Deppman,
Phys. Rev. D \textbf{91}, 054025 (2015)
doi:10.1103/PhysRevD.91.054025
[arXiv:1501.00953 [hep-ph]].

\bibitem{Gao:2017yas}
Y.~Gao, H.~Zheng, L.~L.~Zhu and A.~Bonasera,
Eur. Phys. J. A \textbf{53}, no.10, 197 (2017)
doi:10.1140/epja/i2017-12397-y
[arXiv:1706.03693 [nucl-th]].

\bibitem{Tao:2020uzw}
J.~Q.~Tao, M.~Wang, H.~Zheng, W.~C.~Zhang, L.~L.~Zhu and A.~Bonasera,
[arXiv:2011.05026 [nucl-th]].

\bibitem{Jena:2020wno}
S.~Jena and R.~Gupta,
Phys. Lett. B \textbf{807}, 135551 (2020)
doi:10.1016/j.physletb.2020.135551

\bibitem{Gupta:2020naz}
R.~Gupta, A.~Menon and S.~Jena,
[arXiv:2012.08124 [hep-ph]].

\bibitem{Abbas:2013bpa}
E.~Abbas \textit{et al.} [ALICE],
Phys. Lett. B \textbf{726}, 610-622 (2013)
doi:10.1016/j.physletb.2013.09.022
[arXiv:1304.0347 [nucl-ex]].

\bibitem{Adam:2015kda}
J.~Adam \textit{et al.} [ALICE],
Phys. Lett. B \textbf{754}, 373-385 (2016)
doi:10.1016/j.physletb.2015.12.082
[arXiv:1509.07299 [nucl-ex]].

\bibitem{Acharya:2018hhy}
S.~Acharya \textit{et al.} [ALICE],
Phys. Lett. B \textbf{790}, 35-48 (2019)
doi:10.1016/j.physletb.2018.12.048
[arXiv:1805.04432 [nucl-ex]].

  \bibitem{Pearson343}
 K. Pearson, 
 Philosophical Transactions of the Royal Society of London A: Mathematical, Physical and Engineering Sciences 186, 343 (1895)
 
  \bibitem{pollard}
J. H. Pollard, 
Numerical and Statistical Techniques (Cambridge University Press, 1979)
 
\bibitem{Gupta:2021olm}
R.~Gupta, S.~Jain and S.~Jena,
[arXiv:2103.11185 [hep-ph]].

\bibitem{Tsallis:1987eu} 
  C.~Tsallis,
  \textit{Possible Generalization of Boltzmann-Gibbs Statistics},
  J.\ Statist.\ Phys.\  {\bf 52}, 479 (1988).


\bibitem{Abelev:2012hxa}
B.~Abelev \textit{et al.} [ALICE],
Phys. Lett. B \textbf{720}, 52-62 (2013)
doi:10.1016/j.physletb.2013.01.051
[arXiv:1208.2711 [hep-ex]].

\bibitem{Acharya:2018eaq}
S.~Acharya \textit{et al.} [ALICE],
Phys. Lett. B \textbf{788}, 166-179 (2019)
doi:10.1016/j.physletb.2018.10.052
[arXiv:1805.04399 [nucl-ex]].

\bibitem{root}
 R. Brun and F. Rademakers 1997 ROOT - An Object Oriented Data Analysis Framework,Nucl. Instrum.Meth. A38981. See also ”ROOT [software], Release v6.08.06, doi:10.5281/zenodo.848819
 
\bibitem{James:1975dr}
F.~James and M.~Roos,
\textit{Minuit: A System for Function Minimization and Analysis of the Parameter Errors and Correlations},
Comput. Phys. Commun. \textbf{10}, 343-367 (1975)


\end{thebibliography}
\end{document}